\documentclass[twocolumn]{aastex63}
\usepackage{hyperref,amsmath,amssymb,graphicx}
\numberwithin{equation}{section}
\graphicspath{{Figures/}}

\newcommand {\chandra} {\textit{Chandra}}
\newcommand {\hst} {\textit{HST}}

\begin{document}

\title{X-ray binaries in M51 I: catalog and statistics}

%\correspondingauthor{Jared R. Rice}
\email{jared.r.rice@protonmail.com}

\author[0000-0003-3887-091X]{Jared R. Rice}
\affil{Department of Physics, Texas State University, 601 University Drive, San Marcos, TX 78666, USA}

\author[0000-0002-9282-5207]{Blagoy Rangelov}
\affil{Department of Physics, Texas State University, 601 University Drive, San Marcos, TX 78666, USA}

\author{Andrea Prestwich}
\affil{Center for Astrophysics, Harvard \& Smithsonian, 60 Garden Street, Cambridge, MA 02138, USA}

\author[0000-0003-0085-4623]{Rupali Chandar}
\affil{Department of Physics \& Astronomy, The University of Toledo, Toledo, OH 43606, USA}

\author{Luis Bichon}
\affil{Department of Physics, Texas State University, 601 University Drive, San Marcos, TX 78666, USA}
\affil{Department of Physics \& Astronomy, Vanderbilt University, 2301 Vanderbilt Place, Nashville, TN 37235, USA}

\author{Clint Boldt}
\affil{Department of Physics, Texas State University, 601 University Drive, San Marcos, TX 78666, USA}

\begin{abstract}
We used archival data from the \emph{Chandra X-ray Observatory} (\chandra) and the \emph{Hubble Space Telescope}, to identify 334 candidate X-ray binary (XRB) systems and their potential optical counterparts in the interacting galaxy pair NGC 5194/5195 (M51). We present the catalog and data analysis of X-ray and optical properties for those sources, from the deep $892$ ks \chandra\ observations, along with the magnitudes of candidate optical sources as measured in the $8.16$ ks \hst\ observations. The X-ray luminosity function of the X-ray sources above a few times $10^{36}\, {\rm erg\,s^{-1}}$ follows a power law  $N(>L_{X,b})\propto L_{X,b}^{1-\alpha}$ with $\alpha=1.65\pm0.03$. Aproximately 80\% of sources are variable over a 30 day window. Nearly half of the X-ray sources (173/334) have an optical counterparts within $0\farcs5$.
\end{abstract}

%%%%%%%%%%%%%%%%%%%%%%%%%%%%%%%%%%%%%%%%%%%%%%%%%%%%%%%%%%%%%%%%%%%%%%%%%%%%%%%%
\section{Introduction} \label{sec:intro}

The Whirlpool Galaxy (NGC~5194, M51) and its companion (NGC~5195) are a nearby interacting galaxy pair at a distance of $8.58\pm0.1\text{ Mpc}$ \citep{McQuinn2016} in the constellation Canes Venatici. NGC 5194 is a face-on grand design spiral galaxy that lends itself well to studies of its spiral arms, globular clusters, and X-ray binaries XRBs. There have been a large number of studies of the X-ray sources in M51 going back decades. \cite{Terashima2004} studied the X-ray point source population observed in the two of the earliest M51 {\it Chandra X-Ray Observatory} (\chandra) observations (ObsId 354 \& 1622). In a follow-up, \cite{Terashima2006} investigated the candidate optical counterparts to those X-ray point sources  using \hst\ (with the additional \chandra\ observation ObsId 3932). More recently, \cite{Kuntz2016} used most of the available (at the time) \chandra\ data. 

X-ray binaries (XRBs) are gravitationally bound systems containing a compact object (black hole or neutron star) accreting matter from a main sequence or massive star companion. XRBs fall generally into two main classes: low-mass (LMXBs) and high mass (HMXBs), distinguished by the mass of the companion star. A LMXB has X-ray emission originating in an accretion disk supplied by Roche-lobe overflow of a low-mass ($M_\text{donor}\le 1.0\,M_\odot$) stellar companion. A HMXB has X-ray emission originating in the accretion of the stellar wind of a high-mass ($M_\text{donor}>8.0\,M_\odot$) stellar companion. HMXBs are divided into two sub-classes: those with O-type companions and those with Be companions. Some excellent reviews on XRBs are found in \cite{Shapiro1983}, \cite{Tauris2006}, and \cite{Remillard2006}.

Images of nearby spiral galaxies taken with \chandra\ reveal bright X-ray sources, many of which are believed to be HMXBs. \citet{Fabbiano1989,Fabbiano2006} present an extensive summary of the X-ray source populations in nearby spiral galaxies. X-ray properties such as X-ray luminosity, hardness ratios, and variability can be utilized to identify and study X-ray binaries \citep{Kaaret2001,Prestwich2003,Luan2018,Jin2019,Sell2019}.

\hst\ provides another avenue to investigate XRBs. While the optical magnitudes of LMXBs will be too small to detect, the massive donor stars of HMXBs can be detected with \hst\ at the distance to M51. For example, Supergiant donors are bright, with $M_{V}$ brighter than $\approx-6.5$ \citep{CI1998}, while Be donors tend to be fainter, with typical $M_{V}$ ranging from $-2$ to $-5$ \citep{McBride2008}. Deep photometry with \hst\ can therefore be used to distinguish between the two main classes of HMXBs.

In this paper we present \chandra\ X-ray and \hst\ optical data analysis on the X-ray sources and their stellar counterpart candidates in the M51 system. In \S\ref{sec:observations}, we describe the X-ray and optical observations used in this study, discuss the results in \S\ref{sec:results}, in \S\ref{sec:conclusions}, we describe our conclusions. Due to the large amount of information, here (Paper I) we will primarily show the methodology used to compile our results. We will present a more in-depth analysis of the X-ray source population, as well as analysis of individual sources in a follow up study (Paper II).

%%%%%%%%%%%%%%%%%%%%%%%%%%%%%%%%%%%%%%%%%%%%%%%%%%%%%%%%%%%%%%%%%%%%%%%%%%%%%%%%
\section{Observations and Data Reduction} \label{sec:observations}

%%%%%%%%%%%%%%%%%%%%%%%%%%%%%%%%%%%%%%%%%%%%%%%%%%%%%%%%%%%%%%%%%%%%%%%%%%%%%%%%
\subsection{Chandra X-ray Observatory}\label{subsec:Chandra}

The Whirlpool Galaxy was the focus of many \chandra\ programs since 2000. For this work, we select data with exposure time $t_{\text{exp}} \ge 10.0\text{ ks}$, resulting in 13 \chandra\ observations, the longest of which is 189\,ks. Information about the X-ray data is listed in Table~\ref{tbl-1}. The data were taken with the Advanced CCD Imaging Spectrometer (ACIS) instrument onboard \chandra. The data were analyzed with the Chandra Interactive Analysis of Observations (CIAO) software version $4.10$ and Chandra Calibration Data Base (CALDB) version 4.7.9\footnote{\href{http://cxc.harvard.edu/ciao/}{http://cxc.harvard.edu/ciao/}}

We aligned all datasets with \emph{US Naval Observatory Robotic Astrometric Camera} (USNO URAT1\footnote{\href{https://www.usno.navy.mil/USNO/astrometry/optical-IR-prod/urat}{https://www.usno.navy.mil/USNO/astrometry/optical-IR-prod/urat}}) Catalog using the CIAO scripts \texttt{wcs\_match} and \texttt{wcs\_update}. Taking into account the new aspect ratio solution and bad pixel files, the observation event files were merged into one event file using \texttt{merge\_obs}. The CIAO script \texttt{mkpsfmap} was run on the full merged event file, taking the minimum PSF map size at each pixel location.

We used the CIAO's Mexican-hat wavelet source detection routine \texttt{wavdetect} \citep{Freeman2002} on the merged data to create source lists. Wavelets of 1, 2, 4, 6, 8, 12, 16, 24, and 32 pixels and a detection threshold of $10^{-6}$ were used, which typically results in one spurious detection per million pixels. 

\begin{figure*}
\begin{center}
\includegraphics[width=\linewidth]{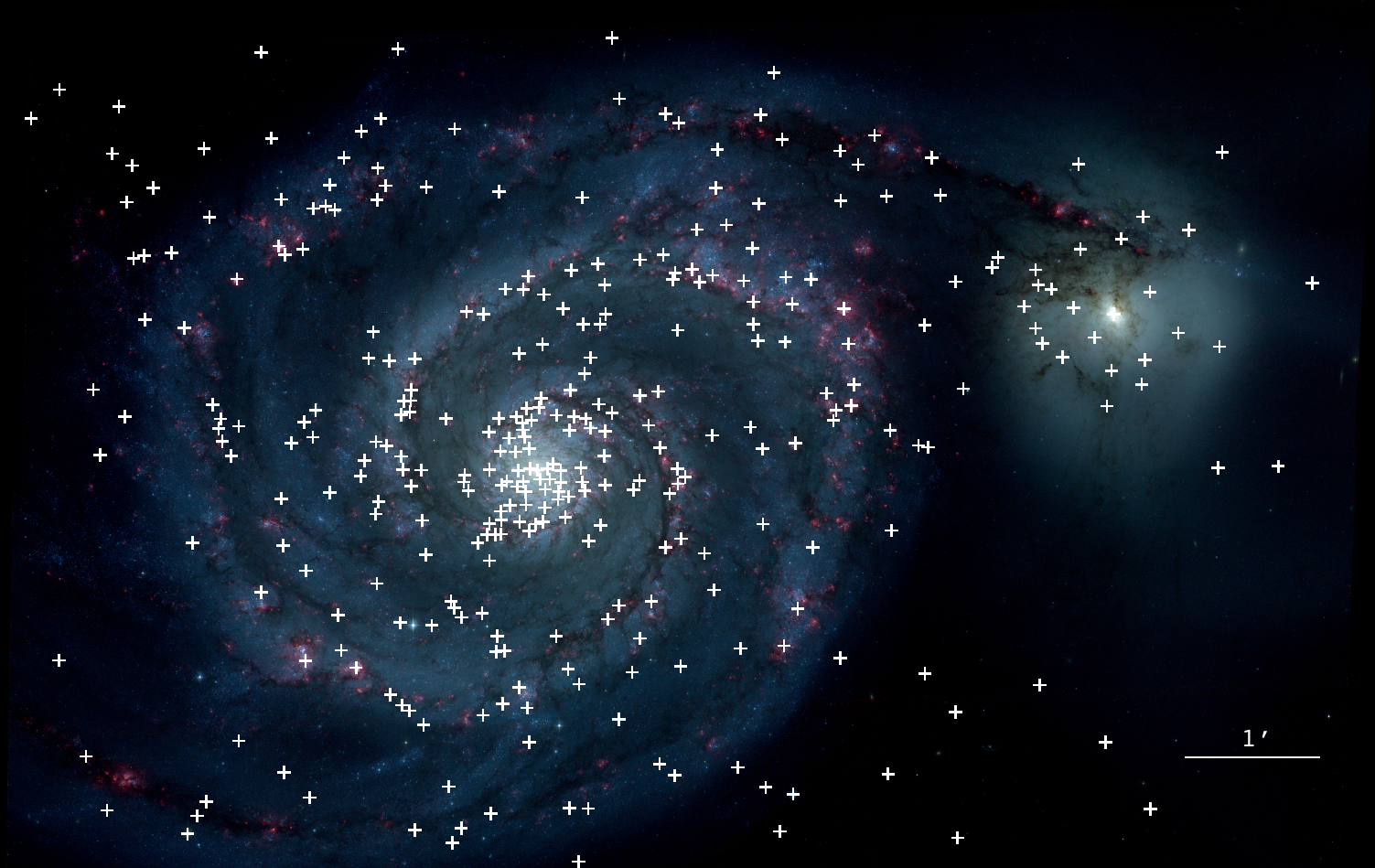}
\caption{\label{fig:m51}Combined \hst\ (r: F658N, g: F555W, b: F435W) image with candidate X-ray point source centroids overlaid as white crosses. North is to the right.}
\end{center}
\end{figure*}

We followed standard CIAO procedures\footnote{\url{http://cxc.harvard.edu/ciao/threads/wavdetect_merged/}}, using an exposure-time-weighted average PSF map in the calculation of the merged PSF. We detected a total of 497 X-ray sources in the merged dataset. In this paper we focus on the sources that are also withing the \hst\ field-of-view, of which there are left 334 (Figure~\ref{fig:m51}). The \texttt{srcflux} CIAO tool was then run individually on each observation (using the coordinates found by \texttt{wavdetect}). The data have been restricted to the energy range between 0.5 and 7.0\,keV and filtered in three energy bands, 0.5--1.2\,keV (soft), 1.2--2.0\,keV (medium), and 2.0--7.0\,keV (hard). We corrected our source catalog to the effects of neutral hydrogen absorption along the line of sight using the Galactic Neutral Hydrogen Density Calculator (COLDEN\footnote{\href{https://cxc.harvard.edu/toolkit/colden.jsp}{https://cxc.harvard.edu/toolkit/colden.jsp}}) tool, finding a mean neutral hydrogen absorption along the line of sight to each source of $n_{\rm {H}} = (1.53\pm0.03)\times10^{20}\text{ cm}^{-2}$. Our fluxes are consistent with the Chandra Source Catalog v2 (CSC\footnote{\url{https://cxc.harvard.edu/csc/}}).

%%%%%%%%%%%%%%%%%%%%%%%%%%%%%%%%%%%%%%%%%%%%%%%%%%%%%%%%%%%%%%%%%%%%%%%%%%%%%%%%
\begin{deluxetable}{cccccc}
%\tabletypesize{\scriptsize}
\tablecaption{\chandra\ Observations\label{tbl-1}}
\tablewidth{0pt}
%\tablecolumns{6}
\tablehead{
\colhead{ObsId} & \colhead{Date} & \colhead{Detector} & \colhead{Mode\tablenotemark{a}} & \colhead{PI} &  
\colhead{Exp\tablenotemark{b}}}
\startdata
354 & 2000-06-20 & ACIS-S & F & Wilson & 15 \\
1622 & 2001-06-23 & ACIS-S & VF & Wilson & 29 \\
3932 & 2003-08-07 & ACIS-S & VF & Terashima & 50 \\
12562 & 2011-06-12 & ACIS-S & VF & Pooley & 10 \\
12668 & 2011-07-03 & ACIS-S & VF & Soderberg & 10 \\
13813 & 2012-09-09 & ACIS-S & F & Kuntz & 180 \\
13812 & 2012-09-12 & ACIS-S & F & Kuntz & 180 \\
15496 & 2012-09-19 & ACIS-S & F & Kuntz & 40 \\
13814 & 2012-09-20 & ACIS-S & F & Kuntz & 190 \\
13815 & 2012-09-23 & ACIS-S & F & Kuntz & 68 \\
13816 & 2012-09-26 & ACIS-S & F & Kuntz & 74 \\
15553 & 2012-10-10 & ACIS-S & F & Kuntz & 38 \\
19522 & 2017-03-17 & ACIS-I & F & Brightman & 40 \\
\enddata
\tablenotetext{a}{F = ``Faint'', VF = ``Very Faint''}
\tablenotetext{b}{Proposed exposure in ks.}
\end{deluxetable}
%%%%%%%%%%%%%%%%%%%%%%%%%%%%%%%%%%%%%%%%%%%%%%%%%%%%%%%%%%%%%%%%%%%%%%%%%%%%%%%%

%%%%%%%%%%%%%%%%%%%%%%%%%%%%%%%%%%%%%%%%%%%%%%%%%%%%%%%%%%%%%%%%%%%%%%%%%%%%%%%%
\subsection{Hubble Space Telescope} \label{subsec:hst}

A six-image mosaic image of M51 with the {\it Hubble Space Telescope} (\hst) Advanced Camera for Surveys was obtained by the Hubble Heritage Team\footnote{\href{https://archive.stsci.edu/prepds/m51/index.html}{https://archive.stsci.edu/prepds/m51/index.html}} (PI: Beckwith, program GO~10452) in January 2005 (see \citealt{Mutchler2005}). The pixel scale of these observations is $0.05''$\,pix$^{-1}$, corresponding to 2.1\,pc\,pix$^{-1}$ at the observed distance of M51. The full mosaic consists of four bands $I$, $V$, $B$, and $H\alpha$ with exposure times of $1360$, $1360$, $2720$, and $2720$ seconds, respectively. The total exposure time is thus $t_{exp} = 8160$\,s over 96 separate exposures. We identified sources in each of the four \hst\ images to align with the URAT1 Catalog and improve the absolute astrometry of the images (similar to \chandra). The common sources totaled 43, distributed across the M51 system. In IRAF, the command \texttt{ccmap} was run on all four of the \hst\ images. The \texttt{ccmap} command finds a six-parameter linear coordinate transformation (plate solution) that takes the $(X,Y)$ centroids and maps them to the more accurate astrometric positions (URAT1 Catalog). In the four bands ($I,V,H\alpha,B$) the mean $(\text{RA},\text{Dec})$ offsets were $(0.142'',0.119'')$, $(0.141'',0.124'')$, $(0.143'',0.124'')$, and $(0.144'',0.117'')$, respectively. We identified candidate \hst\ point sources that fell within $0.5''\,(10\,\text{px})$ of the 334 \chandra\ X-ray point source centroids in our X-ray catalog. We chose $0.5''$ to limit the total number of sources in the catalog while making sure all candidate optical counterparts were identified.

We used the AstroPy package \texttt{photutils}\footnote{\href{https://photutils.readthedocs.io/en/stable/}{https://photutils.readthedocs.io/en/stable/}} to perform photometry calculations on the candidate \hst\ sources. Within \texttt{photutils} we created a circular aperture of radius $r=3.0$ px around each source. The background counts were summed within an annulus centered on each \hst\ point source with inner radius $r_{in}=8.0$\,px and outer radius $r_{out}=11.0$\,px. We corrected for the encircled energy fraction (EEF) using the most recent ACS encircled energy values\footnote{\href{https://www.stsci.edu/hst/instrumentation/acs/data-analysis/aperture-corrections}{https://www.stsci.edu/hst/instrumentation/acs/data-analysis/aperture-corrections}}. The output of \texttt{photutils} on the \hst\ data includes the corrected $(I,V,H\alpha,B)$ magnitudes in the VegaMag system\footnote{\href{https://www.stsci.edu/hst/instrumentation/acs/data-analysis/zeropoints}{https://www.stsci.edu/hst/instrumentation/acs/data-analysis/zeropoints}} for each candidate point source.

%%%%%%%%%%%%%%%%%%%%%%%%%%%%%%%%%%%%%%%%%%%%%%%%%%%%%%%%%%%%%%%%%%%%%%%%%%%%%%%%
%%%%%%%%%%%%%%%%%%%%%%%%%%%%%%%%%%%%%%%%%%%%%%%%%%%%%%%%%%%%%%%%%%%%%%%%%%%%%%%%

%%%%%%%%%%%%%%%%%%%%%%%%%%%%%%%%%%%%%%%%%%%%%%%%%%%%%%%%%%%%%%%%%%%%%%%%%%%%%%%%
\begin{figure*}
\begin{center}
\includegraphics[width=\columnwidth]{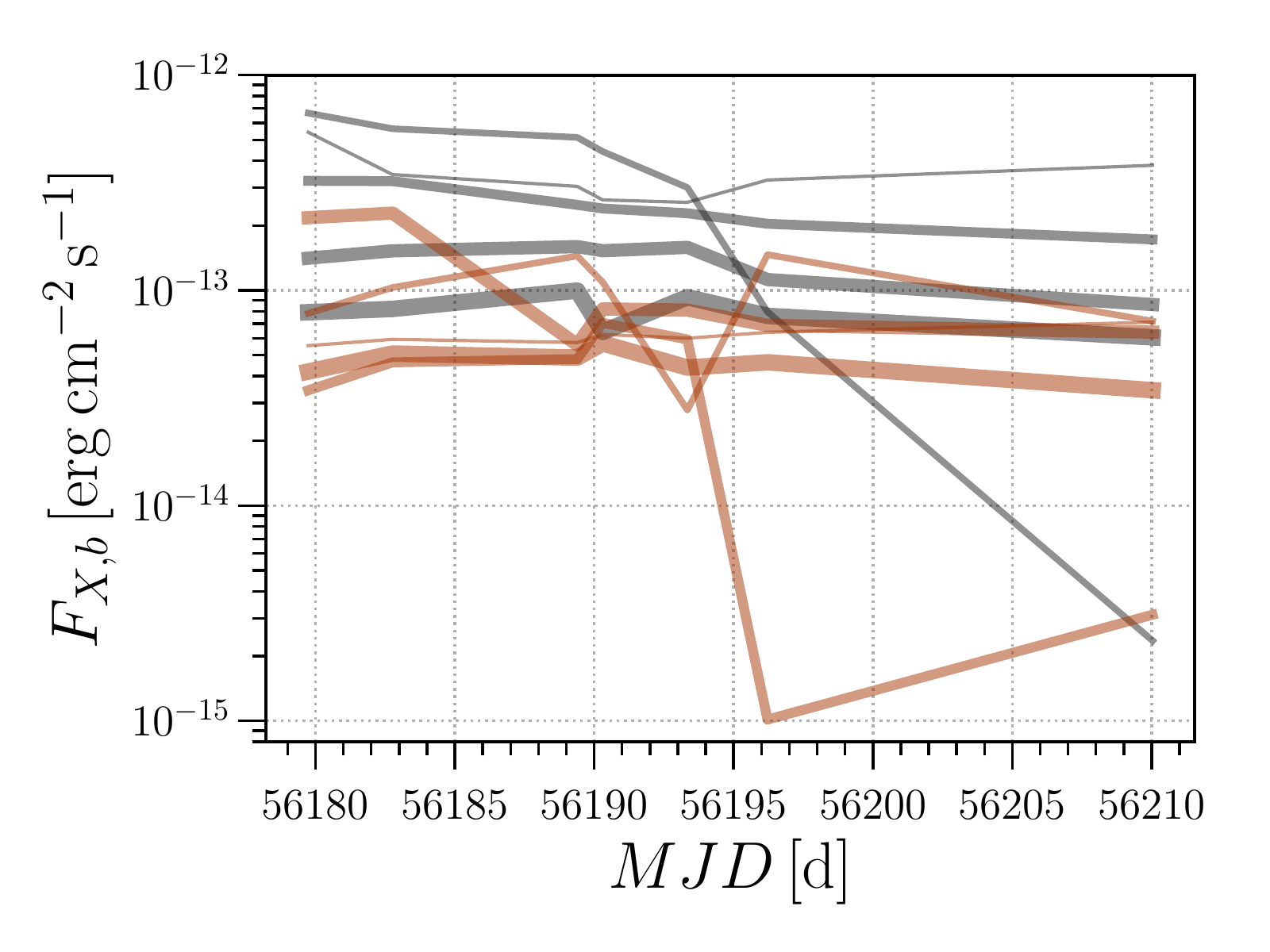}
\includegraphics[width=\columnwidth]{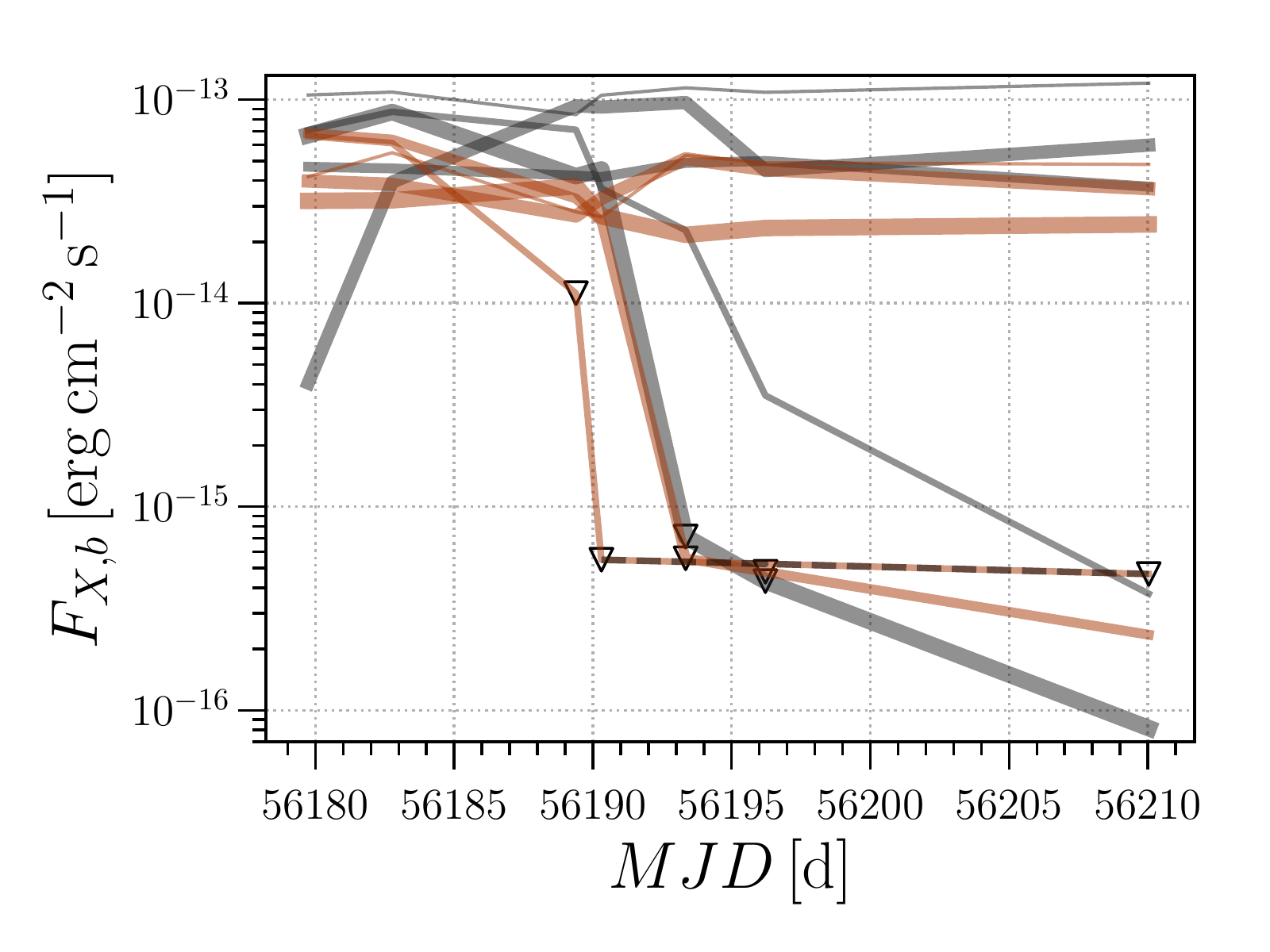}
\caption{\label{fig:fb09} Broad X-ray flux light curves for the top twenty (by net counts) X-ray sources during ObsId 13813, 13812, 15496, 13814, 13815, 13816, and 15553. {\bf Left}: Broad X-ray flux light curves of the brightest 1-5 (black curves) and next brightest 6-10 (copper curves) X-ray sources. {\bf Right}: Broad X-ray flux light curves of the next brightest 10-15 (black curves) and next brightest 16-20 (copper curves) X-ray sources. The black triangles indicate upper limits, while the dashed black and copper line indicates one particular source ($x_{id}=199$: RA: 13:30:06.0397, DEC: +47:15:42.477) was outside the FOV for ObsIDs 13814, 13815, and 13816 and thus had no measured X-ray counts in those observations, except for a 90\% upper limit on the counts during ObsID 13814. In both panels, in the 30-day window of these observations, some sources vary in flux by approximately two orders of magnitude. {\bf Note}: Line thickness increases with decreasing net counts within each group of five curves. Also note the different $y$-axis scales. The time-averaged mean broad flux error of these twenty sources over the 30-day window of observations in this figure is $\langle\delta F_{X,b}\rangle\simeq\,5.7\times10^{-15}\,{\rm erg}\,{\rm cm}^{-2}\,{\rm s}^{-1}$, and for most of the data points the error is within the line thickness.}
\end{center}
\end{figure*}
%%%%%%%%%%%%%%%%%%%%%%%%%%%%%%%%%%%%%%%%%%%%%%%%%%%%%%%%%%%%%%%%%%%%%%%%%%%%%%%%

%%%%%%%%%%%%%%%%%%%%%%%%%%%%%%%%%%%%%%%%%%%%%%%%%%%%%%%%%%%%%%%%%%%%%%%%%%%%%%%%
\subsection{Optical Counterparts}\label{subsec:counterparts}

Candidate point source optical counterparts were found by identifying the brightest \hst\ point source within the \chandra\ positional uncertainty of each X-ray source. We used a 90\% confidence level positional uncertainty of $0.5''$ typical for a $5'$ off-axis X-ray source with 50 counts (see Eq. 12 in \citealt{Kim2007}). This positional uncertainty corresponds to $20.8$\,pc at the distance of M51. In total, there are 173 such candidate optical counterparts. The closest \hst\ source to the X-ray centroid is not always the brightest and often is not visible in all four \hst\ bands, so we justify the optical counterpart candidate identification process in this way. It is possible that the true physical counterparts are invisible in the four \hst\ bands and we identify the incorrect physical counterpart using this method, but it seems to capture the majority of the sources sufficiently. We used the $I$-band images to select the brightest candidate optical counterpart within the \chandra\ $1\sigma$ uncertainty. If we select the closest candidate optical counterpart in the same way, we pick up $\sim 65\%$ (113/173) of the same sources; that is, 60 of the candidate \hst\ counterparts are the brightest, but not closest sources to the X-ray centroids.

%%%%%%%%%%%%%%%%%%%%%%%%%%%%%%%%%%%%%%%%%%%%%%%%%%%%%%%%%%%%%%%%%%%%%%%%%%%%%%%%
\section{Results \& Discussion}\label{sec:results}

%%%%%%%%%%%%%%%%%%%%%%%%%%%%%%%%%%%%%%%%%%%%%%%%%%%%%%%%%%%%%%%%%%%%%%%%%%%%%%%%
\subsection{X-ray Variability}\label{subsec:xrayvariability}

We look for short term X-ray variability using seven \chandra\ observations from 2012 (ObsId 13813, 13812, 15496, 13814, 13815, 13816, and 15553). These observations span over a one month period (see Table~\ref{tbl-1}). In Figure~\ref{fig:fb09}, we plot the broad X-ray flux light curves of the brightest twenty (by net counts) X-ray sources. The brightest ten are shown in the left panel, while the next brightest ten are shown in the right panel. In the 30-day window of these observations, some sources vary in flux by approximately two orders of magnitude.

We calculate a reduced chi-square statistic, $\chi^2_\nu$, for each broad flux light curve in the 30-day window as a measure of variability. We assume the null hypothesis that the underlying broad X-ray flux light curve is described by a uniform function whose value is the weighted mean of the flux across the seven observations in the 30-day window. The chi-square statistic $\chi^2$ used here is defined as
\begin{align}
    \chi^2 &\equiv \sum\limits_{i=1}^\nu \frac{(F_i-\mu_i)^2}{\sigma_i^2},
\end{align}
where $F_i$ is the X-ray flux of the $i$th source, $\mu_i$ is the weighted mean error of the associated $i$th flux measurement, and $\sigma_i^2$ is the variance of the $i$th flux measurement. The reduced chi-square statistic is simply $\chi^2_\nu\equiv\chi^2/\nu$. The mean of the chi-square distribution is $\nu$, so that $\chi^2_\nu = 1$ is a natural value with which to compare results. The value of $\chi^2_\nu$ should be approximately unity if the null hypothesis is to be accepted. Large values of $\chi^2_\nu$ indicate that the null hypothesis should be rejected. Thus, the sources with $\chi^2_\nu \ge 1$ are sources whose flux varies greatly in the 30-day window and we label them ``variable'' sources. Sources with $\chi^2_\nu \lesssim 1$ are sources that have a light curve in the 30-day window that is consistent with the null hypothesis (uniform flux).

Approximately $80\%$ ($266/334$) of the sources are considered variable by our $\chi^2_\nu \ge 1$ criterion. Approximately $69\%$ (120/173) of the sources with at least one detected candidate stellar counterpart and no cluster counterparts are variable (see our upcoming follow-up Paper II for a discussion of cluster counterparts), while about $77\%$ (124/161) of the sources without a stellar or cluster counterpart are variable. In addition, about $76\%$ (22/29) of the X-ray sources that have both an associated candidate stellar source and candidate cluster are considered variable. There is a strong positive correlation between the variability and flux of the X-ray sources. Our findings are consistent with the inter-observation variability reported in the CSC (for sources that overlap, which is the majority of sources), even though we have limited our variability study to data within this 30-day window. We speculate that the observed strong correlation is due to the small uncertainty associated with very bright sources (see Figure~\ref{fig:fb09}), i.e. the time-averaged mean broad flux error over the 30-day window of observations is $\langle\delta F_{X,b}\rangle\simeq\,5.7\times10^{-15}\,{\rm erg}\,{\rm cm}^{-2}\,{\rm s}^{-1}$.

%%%%%%%%%%%%%%%%%%%%%%%%%%%%%%%%%%%%%%%%%%%%%%%%%%%%%%%%%%%%%%%%%%%%%%%%%%%%%%%%
\subsection{X-ray Hardness Ratios}\label{subsec:hardness}

We calculate two X-ray hardness ratios (HRs), ``soft'' and ``hard'' ($HR_1$ and $HR_2$, respectively), for all X-ray sources using the same seven observations as follows:
\begin{align}
    HR_1 &\equiv \frac{M-S}{M+S}\quad\text{and}\\
    HR_2 &\equiv \frac{H-M}{H+M},
\end{align}
\noindent where $S$, $M$, and $H$ are the X-ray counts in each of the \chandra\ bands (soft, medium, and hard) discussed in \S\S\ref{subsec:Chandra}. We also calculate the associated uncertainty in each of the hardness ratios.

\begin{figure}
\begin{center}
\includegraphics[trim= 5mm 5mm 5mm 5mm clip, width=\columnwidth]{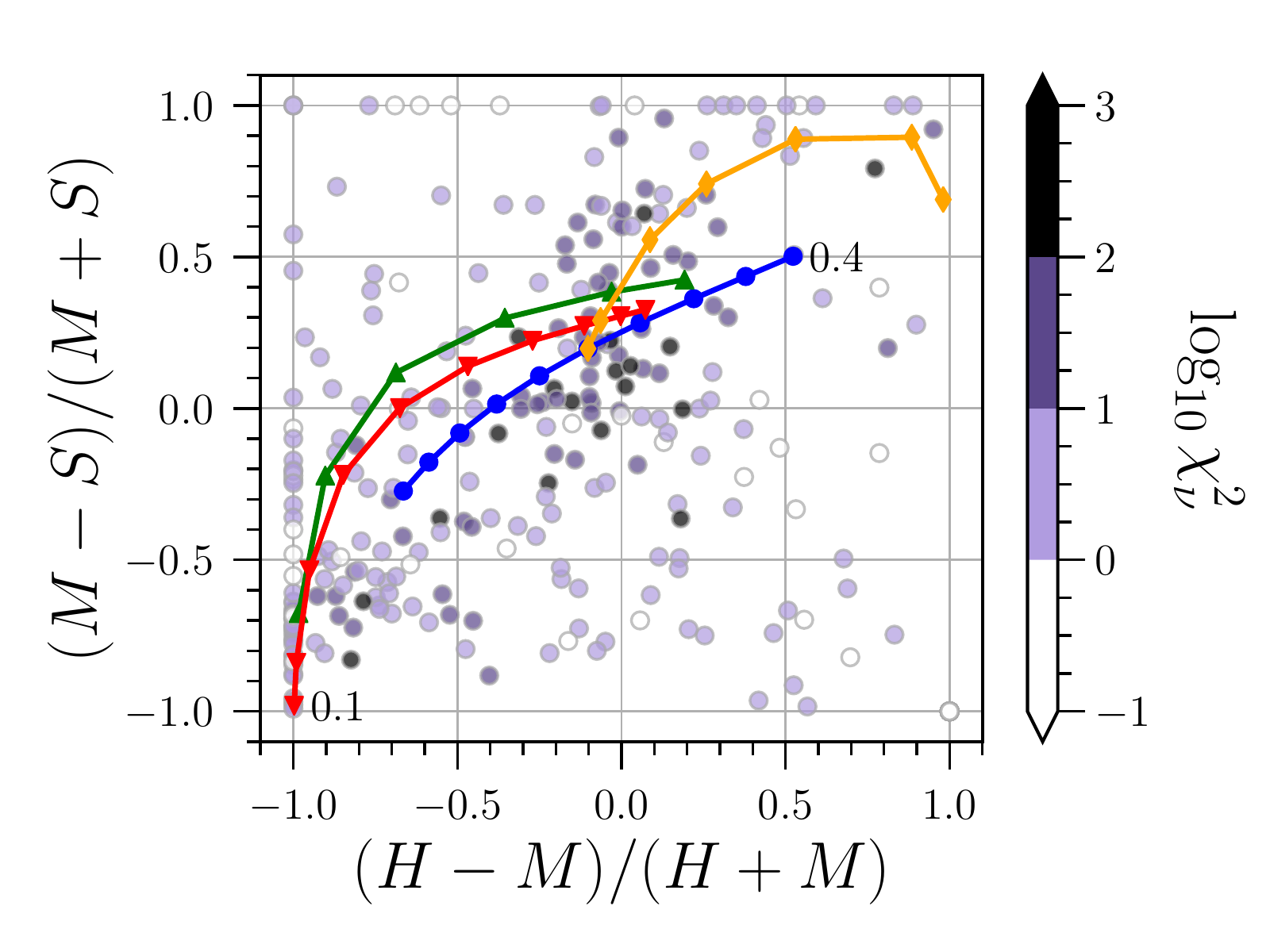}
\includegraphics[trim= 5mm 5mm 5mm 3mm clip, width=\columnwidth]{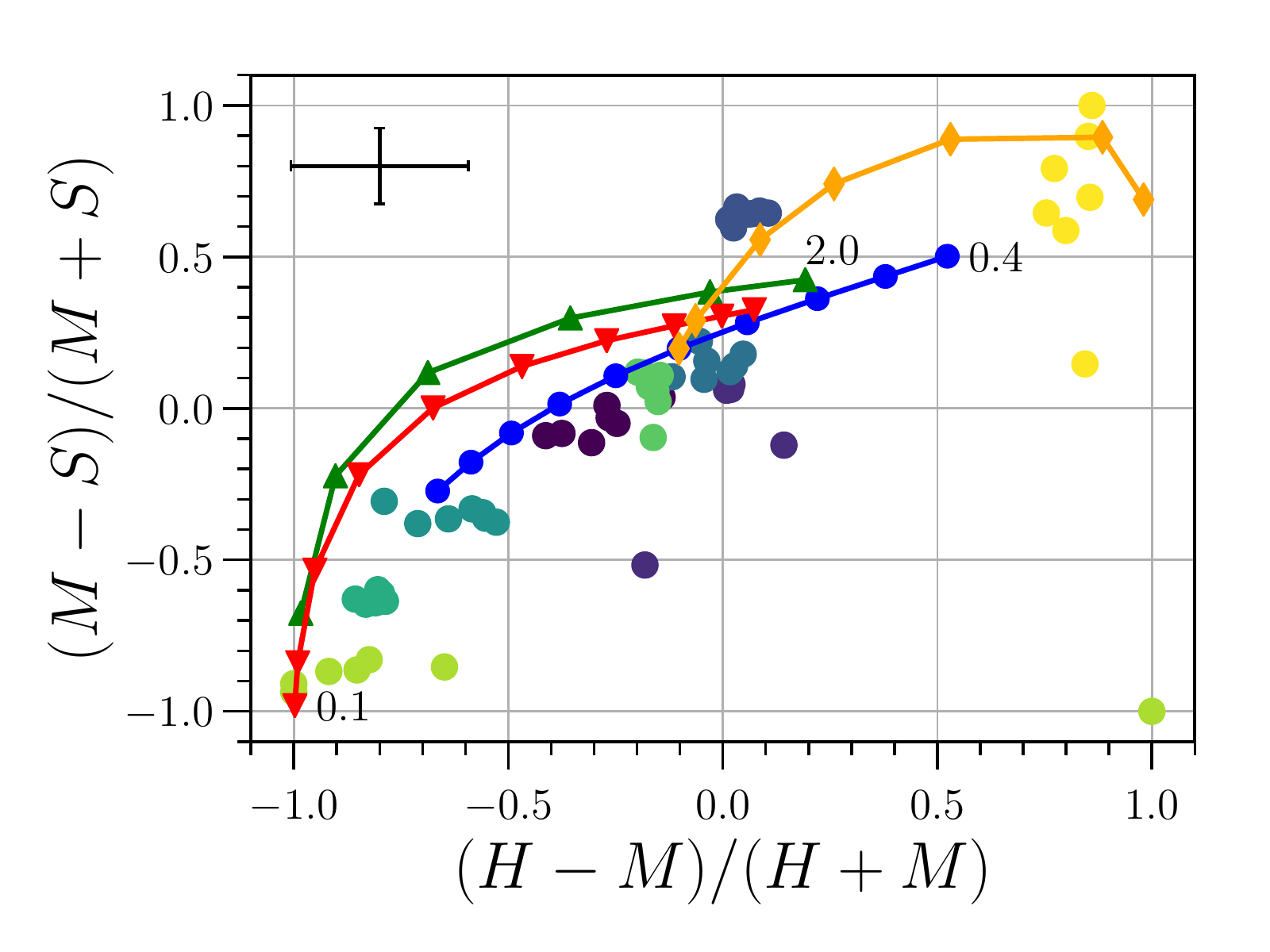}
\includegraphics[trim= 5mm 5mm 5mm 3mm, clip, width=\columnwidth]{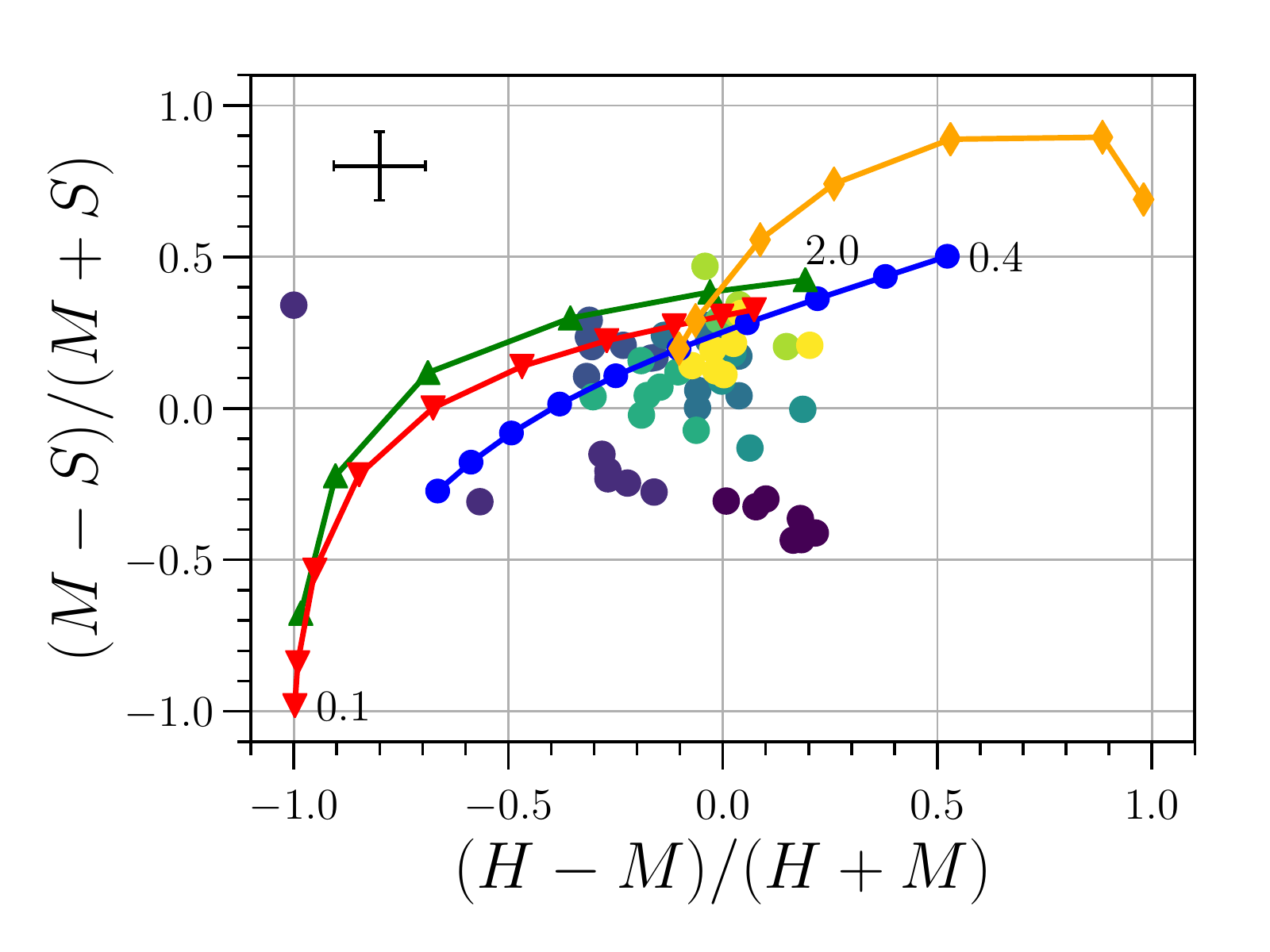}
\caption{\label{fig:hrchi2} {\bf Top}: X-ray color-color diagram for X-ray sources, colored by the logarithm of their reduced chi-square statistic calculated in the 30-day window. The values of $HR_1$ and $HR_2$ in the figure are those of the longest observation (ObsId 13814) in the 30-day window. Overlaid in all three plots are the hardness color evolution tracks of various accretion disk models, see text for details.
{\bf Middle}: Color-color diagram of the brightest ten sources (by net counts), with $1\sigma$ error bar shown in black.
{\bf Bottom}: Color-color diagram of the next brightest ten sources (by net counts), with $1\sigma$ error bar shown in black.
}
\end{center}
\end{figure}

In Figure~\ref{fig:hrchi2}, top panel, we plot the X-ray color-color diagram for all sources, colored by the logarithm of their reduced chi-square statistic calculated in the 30-day window discussed in \S\S\ref{subsec:xrayvariability}. The two X-ray colors are the measurements from the longest of the observations in the 30-day window, ObsId 13814.

Hardness ratio diagrams, such as our Figure~\ref{fig:hrchi2} and Figure~4 in \cite{Prestwich2003} (which uses a different definition\footnote{In \cite{Prestwich2003}, they define the hard and soft X-ray colors as $HR_1\equiv(M-S)/T$ and $HR_2\equiv(H-M)/T$, respectively, where $S$, $M$, $H$, and $T\equiv S+M+H$ are the soft, medium, hard, and total X-ray counts, respectively.} of the X-ray hardness ratios), have been used historically to assist with revealing the nature of the X-ray sources. The majority of the variable sources ($\chi^2_\nu \ge 1$) lie in the XRB (LMXB and HMXB) regions of the figure (see e.g., \citealt{Prestwich2003}), while most of the low variability sources lie in the region of the diagram that is generally occupied by thermal supernova remnants. However, it is well established that X-ray information alone is not enough to accurately identify the nature of unknown X-ray sources. Therefore, we use the X-ray colors together with optical information (see Section~\ref{subsec:opticalcounterparts}) to classify these sources.

In Figure~\ref{fig:hrchi2} we also plot the X-ray color-color diagrams of the brightest twenty (by net counts) X-ray sources; the top ten brightest in the middle panel and the next ten brightest in the bottom panel. Overlaid in all three plots are the hardness color evolution tracks of various accretion disk models. In blue are power law models with increasing photon index $\Gamma$ from $0.4-4$, in orange are absorbed power law models with increasing hydrogen column density, in green are disk blackbody models with temperature ranging from $0.02-2.0\,{\rm keV}$, and in red are absorbed bremsstrahlung models with temperature ranging from $0.1-10.0\,{\rm keV}$. These color-color diagrams contain X-ray colors from all available data in the 30-day window, with appropriate $1\sigma$ error bars. Each source has multiple (same plotted color) points in the diagram, and the color-color evolution is thus apparent. Typically, the color-color evolution is $\lesssim 0.5$ in either color over the entire data set. This suggests that while some spectral change may occur over the 30-day period, the accretion process for these sources does not change dramatically. There are a few sources that appear to significantly change their spectral properties as indicated by movement in the plane of the X-ray color-color diagram, for example the middle panel of Figure~\ref{fig:hrchi2}. It is possible, however, that the movement in the color-color diagram could arise due to a drop in flux, which would raise signficant uncertainties in the location of a source in the diagram due to low count statistics. The error bars are large enough for many of these sources that the spectral evolution cannot be confidently confirmed. Attempting to track the spectral evolution of fainter sources becomes meaningless due to the large uncertainties associated with the X-ray hardness ratio measurements. Detailed analysis of bright sources will be presented in Paper II.

%%%%%%%%%%%%%%%%%%%%%%%%%%%%%%%%%%%%%%%%%%%%%%%%%%%%%%%%%%%%%%%%%%%%%%%%%%%%%%%%
%%%%%%%%%%%%%%%%%%%%%%%%%%%%%%%%%%%%%%%%%%%%%%%%%%%%%%%%%%%%%%%%%%%%%%%%%%%%%%%%

%%%%%%%%%%%%%%%%%%%%%%%%%%%%%%%%%%%%%%%%%%%%%%%%%%%%%%%%%%%%%%%%%%%%%%%%%%%%%%%%
\subsection{X-ray Luminosity Function}\label{subsec:xrayluminosity}

The X-ray luminosity function (XLF) for all the 334 X-ray point source candidates in M51 can be approximated as a power law within a certain luminosity range. We use the differential luminosity function defined as
\begin{align}
    \frac{dN(>L_0)}{dL} &= A\biggl(\frac{L}{L_0}\biggr)^{-\alpha},\label{eq:dXLF}
\end{align}
\noindent where the luminosity $L_0$ is an arbitrary lower limit and $A$ is some normalization constant. Integrating Eq.~\ref{eq:dXLF} gives the luminosity function within a particular range:
\begin{align}
    N(>L_0) &= \frac{AL_0}{1-\alpha}\biggl(\frac{L}{L_0}\biggr)^{1-\alpha},
\end{align}
and the fractional luminosity function is given by
\begin{align}
    f(>L_0) &= \frac{AL_0}{N_{tot}(1-\alpha)}\biggl(\frac{L}{L_0}\biggr)^{1-\alpha},
\end{align}
where $f(>L_0)$ is the fraction of sources with $L>L_0$ and $N_{tot}$ is the total number of sources.

An important luminosity is the Eddington luminosity of a $1.4\,M_\odot$ compact object (the typical mass of NSs) accreting at the Eddington rate:
\begin{align}
    \nonumber\dot{L}_E &\equiv \dot{M}_Ec^2 = \nonumber\frac{4\pi Gm_pc}{\sigma_T}M\\
&\simeq 1.76\times10^{38}\text{ erg$\,$s$^{-1}$}\,\biggl(\frac{M}{1.4\,M_\odot}\biggr),\label{eq:Eddi}
\end{align}
\noindent where $\dot{M}_E$ is the Eddington accretion rate, $M$ is the mass of the accretor, $G$ is Newton's gravitation constant, $m_p$ is the proton mass, $c$ is the speed of light, and $\sigma_T$ is the Thomson scattering cross section for electrons.

\begin{figure*}
\begin{center}
\includegraphics[width=\columnwidth]{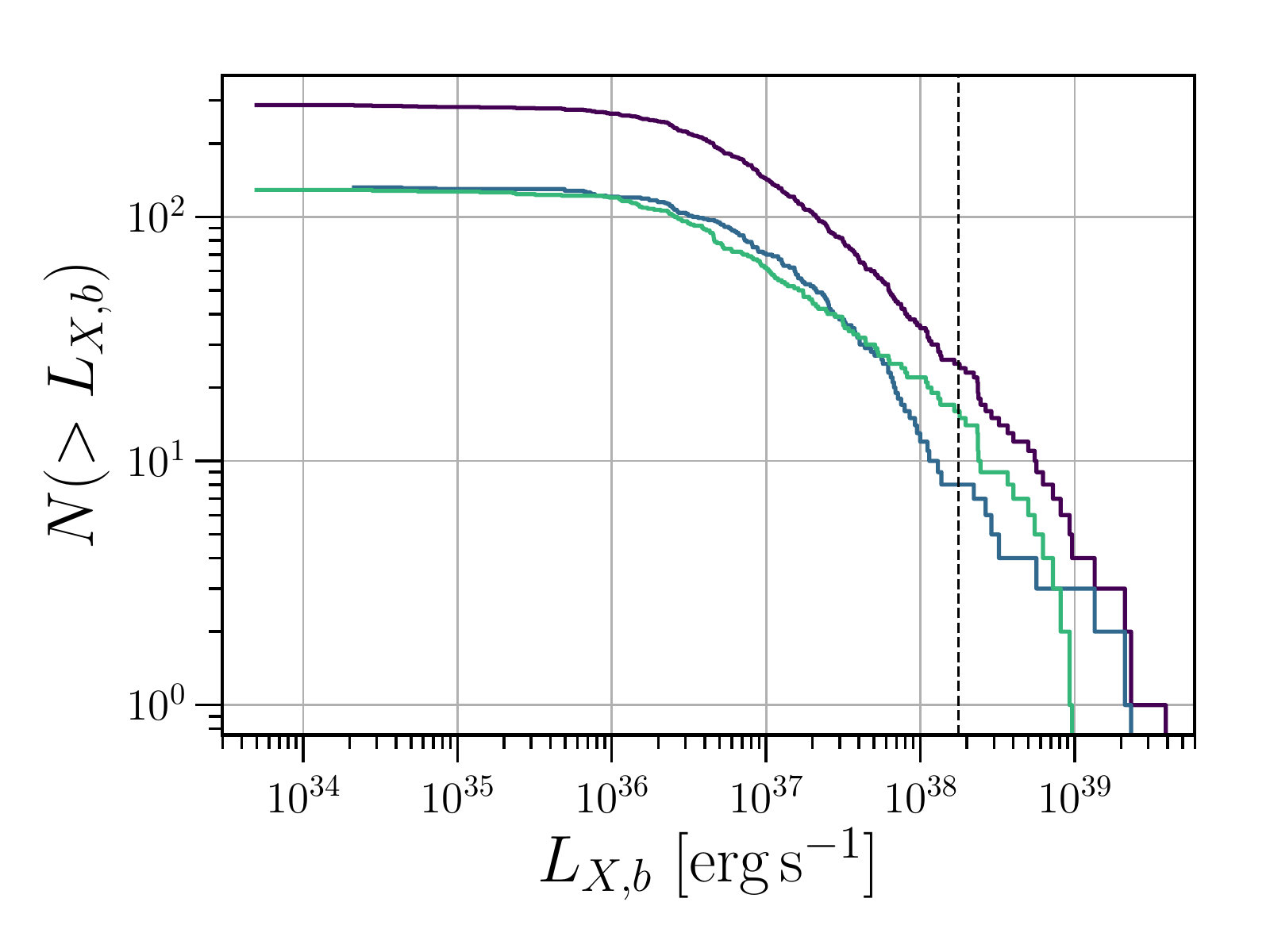}
\includegraphics[width=\columnwidth]{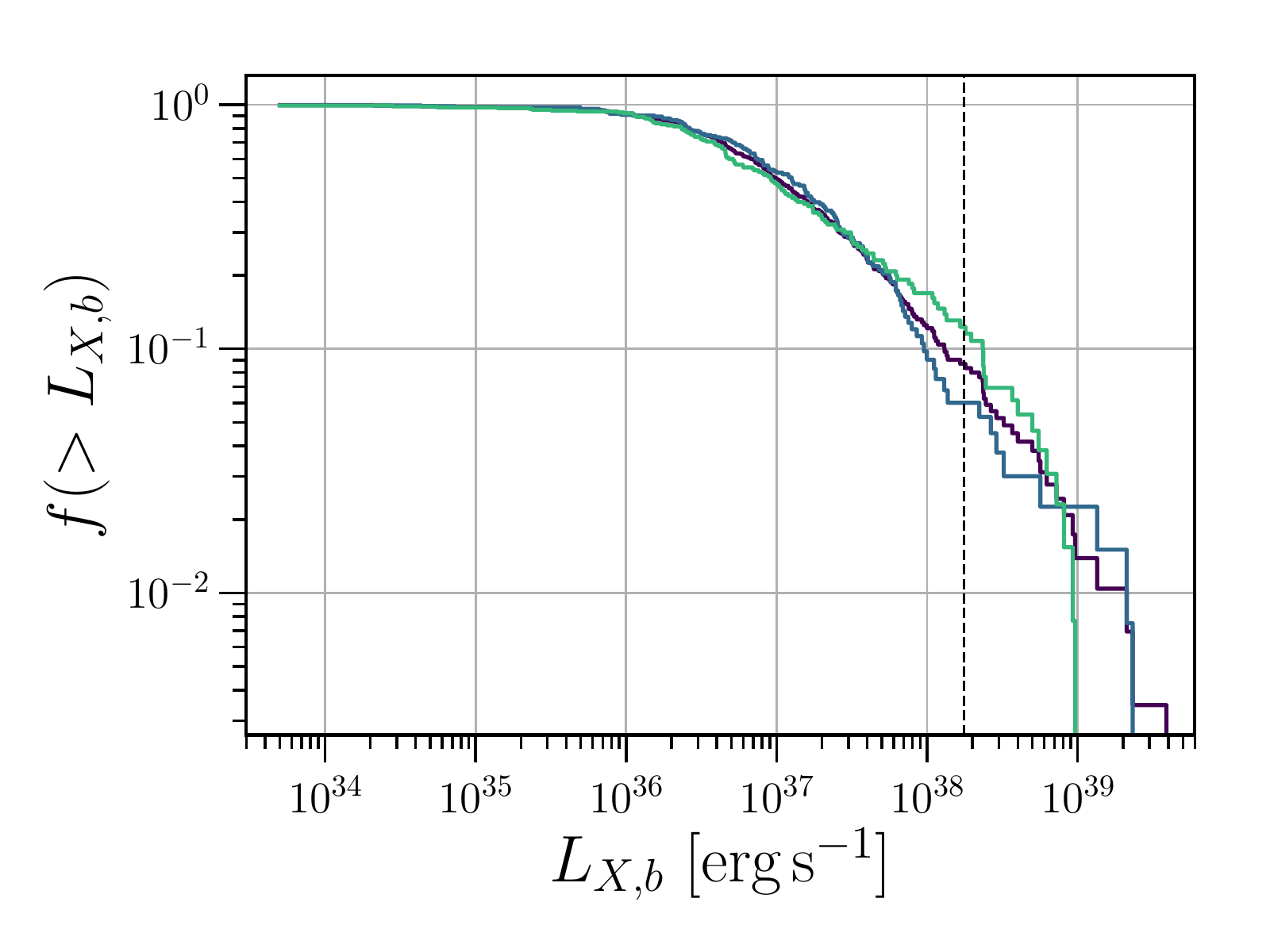}
\caption{\label{fig:XLF} {\bf Left}: Combined X-ray luminosity functions. The purple curve is the full sample of 288/334 X-ray sources that have a measured X-ray luminosity in ObsId 13814. For $L_{X,b}\ge 2\times10^{36}\text{ erg$\,$s$^{-1}$}$, the purple curve follows a power law $N(>L_{X,b})\propto L_{X,b}^{1-\alpha}$ where $\alpha = 1.65\pm0.03$; see text for details. The green curve is a cut of the full sample with 130/334 X-ray sources that have only a stellar \hst\ candidate counterpart within 10 px. The blue curve is a cut of the full sample with 133/334 X-ray sources that have no stellar or cluster counterparts within 10 px. The vertical dashed black line is the Eddington luminosity of a $1.4\,M_\odot$ accretor, i.e. $L_{Edd}\simeq 1.76\times10^{38}\text{ erg$\,$s$^{-1}$}$ (see Eq. \ref{eq:Eddi} in the text). 
{\bf Right}: Combined fractional X-ray luminosity functions. Same as the left panel but normalized.
}
\end{center}
\end{figure*}

In Figure~\ref{fig:XLF}, we plot the combined XLF (total and fractional) on various cuts of the data. The purple curve is the full sample of the $86\%$ (288/334) of X-ray sources that have a measured X-ray luminosity in ObsId 13814 (the observation with the longest exposure time). The green curve is the $39\%$ (130/334) of X-ray sources that have a stellar counterpart in \hst\ (within 10\,px). Across few orders of magnitude of X-ray luminosity starting at $L_{X,b} \ge 2\times10^{36}\text{ erg$\,$s$^{-1}$}$ the curves follow a power law $N(>L_{X,b})\propto L_{X,b}^{-0.65}$, i.e. we fit a power-law to the differential luminosity function with $\alpha = 1.65\pm0.03$. This is consistent with XLFs for star forming galaxies dominated by HMXBs, for example \cite{Lehmer2019} who find $\alpha = 1.59\pm0.05$ for M51. The blue curve represents the X-ray sources that have no stellar or cluster counterparts within 10\,px, $40\%$ (133/334) and has the  same slope. The black vertical dashed line indicates the Eddington luminosity of a canonical NS (e.g., $1.4\,M_\odot$) accretor of $L_{Edd}\simeq 1.8\times10^{38}\text{ erg$\,$s$^{-1}$}$. Fewer than $10\%$ of the sources have an X-ray luminosity that is greater than $L_{Edd}$ for the typical NS accretor.

A major obstacle in studying the extragalactic XRB population is differentiating HMXBs from LMXBs, which cannot be done by their X-ray properties alone. One attempt to solve this problem was done by \cite{Mineo2012} who used galactocentric distance to distinguish between the two types of XRBs. However, many galaxies, including spirals such as M51, show a spatially mixed population of ``young'' and ``old'' XRBs. Our results show that combining \chandra\ and \hst\ data can break this degeneracy.

%%%%%%%%%%%%%%%%%%%%%%%%%%%%%%%%%%%%%%%%%%%%%%%%%%%%%%%%%%%%%%%%%%%%%%%%%%%%%%%%
%%%%%%%%%%%%%%%%%%%%%%%%%%%%%%%%%%%%%%%%%%%%%%%%%%%%%%%%%%%%%%%%%%%%%%%%%%%%%%%%

%%%%%%%%%%%%%%%%%%%%%%%%%%%%%%%%%%%%%%%%%%%%%%%%%%%%%%%%%%%%%%%%%%%%%%%%%%%%%%%%
\subsection{Optical Counterparts to X-ray Sources}\label{subsec:opticalcounterparts}

Due to the distance to M51 there are issues with crowding and source confusion. Many X-rays sources ($51\%$; 173/334) have at least one \hst\ stellar counterpart within the $0\farcs5$ \chandra\ positional uncertainty, whereas ($75\%$; 252/334) have at least one \hst\ stellar counterpart within the 2$\sigma$ \chandra\ uncertainty. Just over half, $51\%$ ($88/173$), of sources that have at least one detected \hst\ stellar counterpart within 1$\sigma$ have at least two detected candidate stellar counterparts.

\begin{figure*}%[h]
\begin{center}
\includegraphics[width=85mm]{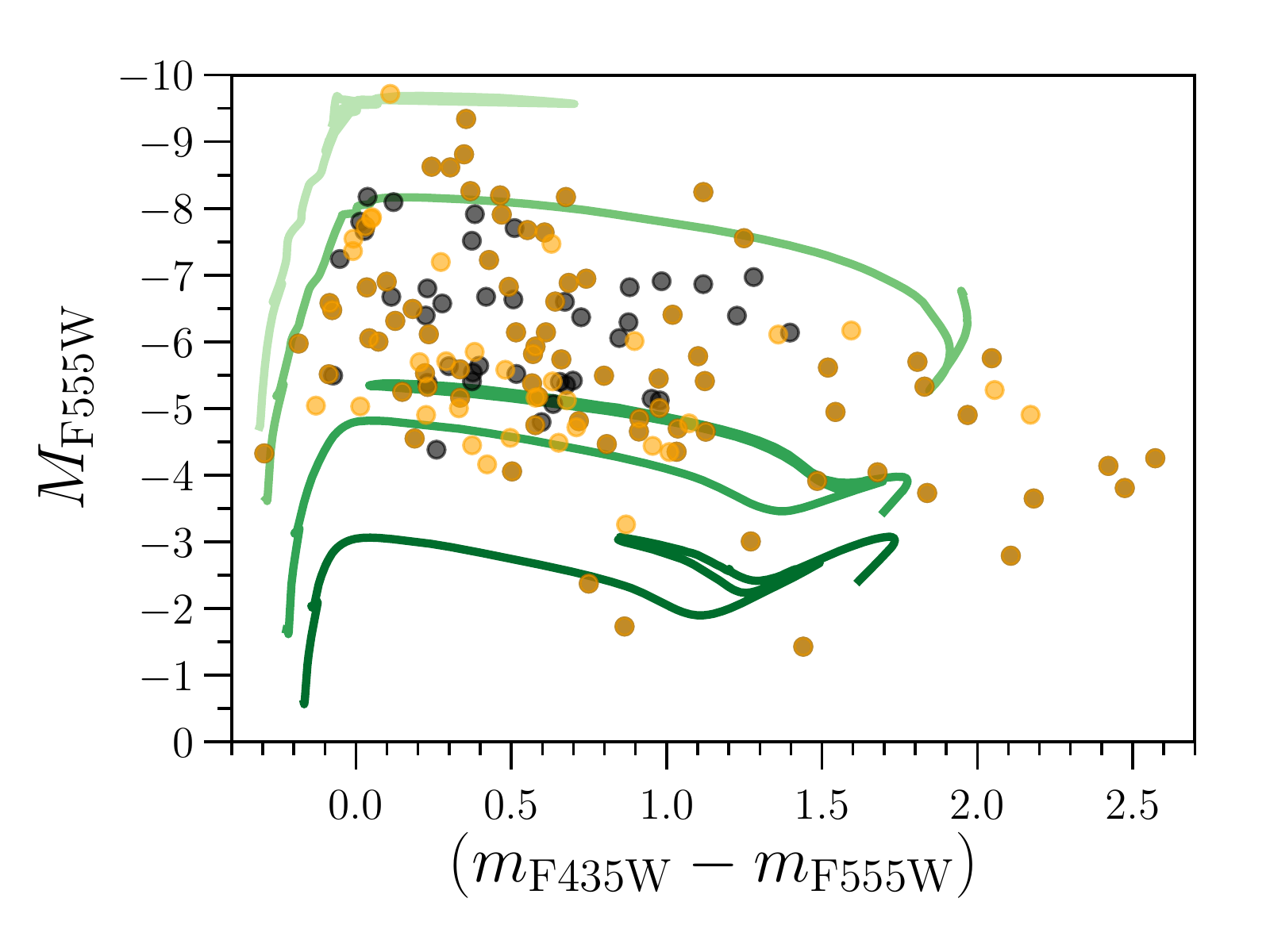}
\includegraphics[width=85mm]{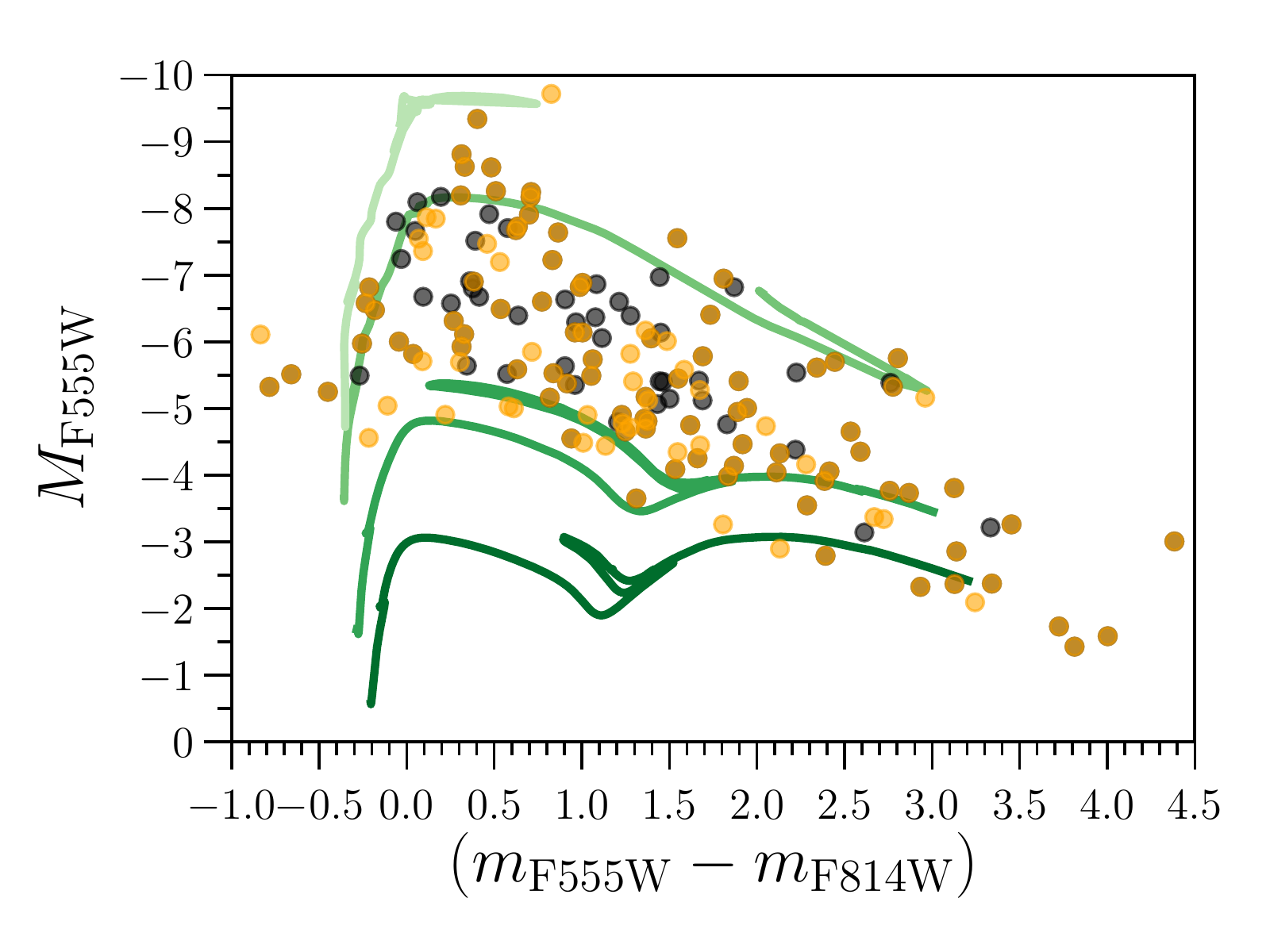}
\caption{\label{fig:b-v_v-i}Color-magnitude diagrams for the potential HST optical counterpart candidates that are the brightest (black) or closest (light orange) within 10 px of the X-ray source centroids. The dark orange indicates a candidate optical counterpart is both the brightest and closest to the X-ray centroid. There are 173 candidate optical counterparts within 10 px of the 334 \emph{Chandra} X-ray point sources. For both panels, the four mass tracks from bottom to top are from MESA Isochrones \& Stellar Tracks (MIST; see text) and have masses $5\,M_\odot$, $8\,M_\odot$, $20\,M_\odot$, and $40\,M_\odot$, respectively. {\bf Left}: $B-V$ color-magnitude diagram. {\bf Right}: $V-I$ color-magnitude diagram.}
\end{center}
\end{figure*}

Selecting the counterpart candidate can be challenging in cases where there are more than two or more optical sources in the search radius. One method of choosing the donor star candidate is to select the closest optical source to the \chandra\ position. On the  other hand, large fraction of the XRBs in M51 are expected to be HMXBs with early-type stars as the donors. Therefore, an alternative method of selecting an optical counterpart is to select the brightest optical sources within the 1$\sigma$ radius ($10$ px). 
%The latter method is illustrated in Figure~\ref{fig:b-v_v-i}. 
In Figure~\ref{fig:b-v_v-i} we plot the $B-V$ and $V-I$ color-magnitude diagrams for the candidate \hst\ optical counterparts that are the brightest or closest within $10$ px of the X-ray point source centroids. If we select the closest candidate \hst\ optical counterpart within $10$ px, out of 173 total candidate optical counterparts, $\sim65.3\%$ (113/173) of the sources are the same. That is, 113 of the \hst\ counterparts are both the brightest and the closest source within $10$ px. As expected, selecting the closest optical counterpart to the X-ray sources is biased toward fainter (and older) stellar sources. However, we performed a two-sample Kolmogorov-Smirnov (K-S) test on the following data from Figure~\ref{fig:b-v_v-i}:
\begin{enumerate}
\item $M_{V,closest}$ vs. $M_{V,brightest}$ ($y$-axis of both panels)
\item $(m_B-m_V)_{closest}$ vs. $(m_B-m_V)_{brightest}$ ($x$-axis of left panel)
\item $(m_V-m_I)_{closest}$ vs. $(m_V-m_I)_{brightest}$ ($x$-axis of right panel),
\end{enumerate}
and found that in each case the null hypothesis $H_0$, namely that the two samples in 1, 2, and 3 above are drawn from the same unknown underlying continuous distribution, cannot be rejected. The two-sample KS test statistic $D$ and $p$-values for each of the three tests above are:
\begin{enumerate}
\item $D = 0.13194$ and $p = 0.15056$
\item $D = 0.07639$ and $p = 0.77899$
\item $D = 0.06250$ and $p = 0.93376$.
\end{enumerate}
At a level of significance $\alpha=0.05$, we cannot reject $H_0$ since in each case $p\ge\alpha$. Thus we cannot claim a statistically significant difference in choosing either the closest or the brightest sources as the candidate optical counterpart to our X-ray sources. The mean photometric error is approximately $0.1$ mag in $V$ and $I$. In Table~\ref{tbl-2}, we select the brightest source as the donor star candidate in case of multiple matches.

Also in Figure~\ref{fig:b-v_v-i} we plot four mass tracks: $5\,M_\odot$, $8\,M_\odot$, $20\,M_\odot$, and $40\,M_\odot$, respectively from bottom to top, taken from the MESA Isochrones \& Stellar Tracks\footnote{\href{http://waps.cfa.harvard.edu/MIST/index.html}{http://waps.cfa.harvard.edu/MIST/index.html}} (see \citealt{Dotter2016,Choi2016,Paxton2011,Paxton2013,Paxton2015,Paxton2018}). The initial protosolar bulk metallicity for the models used is $Z_i = 0.0147$, with extinction $A_V=0$ ($R_V=3.1$). It is clear from the color-magnitude diagram that most of the candidate \hst\ optical counterparts lie above the $8\,M_\odot$ mass track, indicating that most of our candidate sources are likely HMXBs. In classifying the candidate sources as HMXBs, there are no statistically significant differences in choosing either the brightest (black) or closest (light orange) sources (see the two-sample K-S test discussion above).

%%%%%%%%%%%%%%%%%%%%%%%%%%%%%%%%%%%%%%%%%%%%%%%%%%%%%%%%%%%%%%%%%%%%%%%%%%%%%%%%
%%%%%%%%%%%%%%%%%%%%%%%%%%%%%%%%%%%%%%%%%%%%%%%%%%%%%%%%%%%%%%%%%%%%%%%%%%%%%%%%

\section{Conclusions}\label{sec:conclusions}

In this study we presented a catalog and statistical analysis of archival \chandra\ and \hst\ data of point sources in the interacting galaxy pair NGC 5194/5195 (M51). 

\begin{itemize}

\item Using standard CIAO procedures, we detected 334 X-ray point sources in the merged thirteen \chandra\ observations. We corrected the data for neutral hydrogen absorption along the line of sight and improved the astrometry using the USNO URAT1 catalog.

\item We identified 173 candidate optical counterparts to the X-ray sources in our catalog by finding the brightest \hst\ point sources within 10 px of the X-ray source. We found no statistically different results by choosing the closest \hst\ point sources by performing a two-sample Kolmogorov-Smirnov test (see text for details). Similar to \chandra\ the astrometry of the data was corrected by using the USNO URAT1 catalog and applying a six-parameter plate transformation.

\item We calculated a reduced chi-square statistic, $\chi_\nu$, as a measurement of the broad flux variability in a 30-day window of the longest seven observations for the X-ray sources in our catalog and found that approximately $80\%$ of the sources are considered variable, i.e. $\chi_\nu\ge 1$.

\item Approximately $69\%$ of the sources with at least one detected candidate stellar counterpart (but no cluster counterpart) are considered variable and about $77\%$ of the sources without a stellar counterpart are variable (see our upcoming follow-up Paper II for a discussion of candidate cluster counterparts to our X-ray sources).

\item The majority of optical counterparts are above the 8\,M$_\odot$ line in  Figure~\ref{fig:b-v_v-i}, which is consistent with these sources being HMXB candidates.

\item There is a strong positive correlation between the broad X-ray flux and the variability of the X-ray sources in the 30-day window, consistent with the interobservation variability in the CSC catalog.

\item We calculated X-ray hardness ratios for all sources and found that the majority of the variable sources lie in the XRB region of the X-ray color-color diagram (e.g., hard or absorbed X-ray sources; see Figure~\ref{fig:hrchi2}).

\item The broad X-ray luminosity function above a few times $10^{36}\, {\rm erg\,s^{-1}}$ follows a power law  $N(>L_{X,b})\propto L_{X,b}^{1-\alpha}$ with $\alpha=1.65\pm0.03$, consistent with X-ray luminosity functions of star-forming galaxies dominated by HMXBs.

\item Most of the brightest 20 sources do not show any evidence of of flux variability.

\item Fewer than $10\%$ of the X-ray sources have a broad X-ray luminosity greater than the Eddington luminosity of a typical NS accretor.
\end{itemize}

As mentioned earlier, a detailed analysis of individual sources will be presented in a follow-up paper.

\software{CIAO \citep{Fruscione2006}, IRAF \citep{Tody1986,Tody1993}, SAOImageDS9 \citep{Joye2003}, NumPy \citep{Harris2020}, SciPy \citep{Virtanen2020}, Matplotlib \citep{Perez2007}, Astropy \citep{Astropy2013,Astropy2018}}

%%%%%%%%%%%%%%%%%%%%%%%%%%%%%%%%%%%%%%%%%%%%%%%%%%%%%%%%%%%%%%%%%%%%%%%%%%%%%%%%
\section{Acknowledgments}\label{sec:acknowledgments}

We thank an anonymous referee for constructive comments.

%%%%%%%%%%%%%%%%%%%%%%%%%%%%%%%%%%%%%%%%%%%%%%%%%%%%%%%%%%%%%%%%%%%%%%%%%%%%%%%%
\begin{deluxetable*}{cccccccccrr}
\tablecaption{Catalog of X-ray sources\label{tbl-2}}
\tablewidth{0pt}
\tablehead{
\colhead{$X_\text{ID}$} & \colhead{RA} & \colhead{Dec} & \colhead{$C_\text{net}/10^3$} & \colhead{$F_{X,b}/(10^{-15}\text{erg}\,\text{s}^{-1})$} &  
\colhead{$I$} & \colhead{$V$} & \colhead{$H\alpha$} & \colhead{$B$} & \colhead{$HR_1$} & \colhead{$HR_2$}}
\startdata
$178$ & $202.504274$ & $47.228885$ &  $6.73$ & $441$ & $-8.26$ & $-7.58$ & $-8.08$ & $-6.62$ & $0.073$ & $0.012$\\
$62$ & $202.531533$ & $47.185057$ & $6.03$ & $263$ & $-6.70$ & $-5.10$ & $-7.27$ & $-4.69$ & $-0.083$ & $-0.37$\\
$166$ & $202.496180$ & $47.221801$ & $3.63$ & $240$ & $-6.18$ & $-3.83$ & $-4.91$ & $-2.62$ & $0.64$ & $0.069$\\
$190$ & $202.473939$ & $47.243281$ & $2.25$ & $153$ & $-$ & $-$ & $-$ & $-$ & $0.14$ & $0.027$\\
$154$ & $202.414552$ & $47.212141$ & $2.05$ & $109$ & $-6.35$ & $-6.58$ & $-6.34$ & $-6.67$ & $0.023$ & $-0.15$\\
$100$ & $202.470043$ & $47.194362$ & $1.90$ & $62.4$ & $-8.49$& $-7.47$ & $-8.05$ & $-6.49$ & $-0.64$ & $-0.79$\\
$94$ & $202.430548$ & $47.193022$ & $1.77$ & $70.7$ & $-6.97$ & $-$ & $-$ & $-$ & $-0.83$ & $-0.82$\\
$149$ & $202.416637$ & $47.210268$ & $1.65$ & $64.1$ & $-5.72$ & $-5.74$ & $-6.14$ & $-5.67$ & $-0.36$ & $-0.55$\\
$168$ & $202.517998$ & $47.222471$	& $1.38$ & $92.0$ & $-8.06$ & $-7.23$ & $-8.15$ & $-6.80$ & $0.22$ & $-0.033$\\	
$54$ & $202.489953$ & $47.180183$ & $1.08$ & $56.7$ & $-7.11$ & $-5.65$ & $-5.86$ & $-5.42$ & $0.065$ & $-0.21$
\enddata
\end{deluxetable*}
%%%%%%%%%%%%%%%%%%%%%%%%%%%%%%%%%%%%%%%%%%%%%%%%%%%%%%%%%%%%%%%%%%%%%%%%%%%%%%%%

%%%%%%%%%%%%%%%%%%%%%%%%%%%%%%%%%%%%%%%%%%%%%%%%%%%%%%%%%%%%%%%%%%%%%%%%%%%%%%%%

\newpage
\bibliography{references}{}

\begin{thebibliography}{}
\expandafter\ifx\csname natexlab\endcsname\relax\def\natexlab#1{#1}\fi
\providecommand{\url}[1]{\href{#1}{#1}}
\providecommand{\dodoi}[1]{doi:~\href{http://doi.org/#1}{\nolinkurl{#1}}}
\providecommand{\doeprint}[1]{\href{http://ascl.net/#1}{\nolinkurl{http://ascl.net/#1}}}
\providecommand{\doarXiv}[1]{\href{https://arxiv.org/abs/#1}{\nolinkurl{https://arxiv.org/abs/#1}}}

\bibitem[{{Astropy Collaboration} {et~al.}(2013){Astropy Collaboration},
  {Robitaille}, {Tollerud}, {Greenfield}, {Droettboom}, {Bray}, {Aldcroft},
  {Davis}, {Ginsburg}, {Price-Whelan}, {Kerzendorf}, {Conley}, {Crighton},
  {Barbary}, {Muna}, {Ferguson}, {Grollier}, {Parikh}, {Nair}, {Unther},
  {Deil}, {Woillez}, {Conseil}, {Kramer}, {Turner}, {Singer}, {Fox}, {Weaver},
  {Zabalza}, {Edwards}, {Azalee Bostroem}, {Burke}, {Casey}, {Crawford},
  {Dencheva}, {Ely}, {Jenness}, {Labrie}, {Lim}, {Pierfederici}, {Pontzen},
  {Ptak}, {Refsdal}, {Servillat}, \& {Streicher}}]{Astropy2013}
{Astropy Collaboration}, {Robitaille}, T.~P., {Tollerud}, E.~J., {et~al.} 2013,
  \aap, 558, A33, \dodoi{10.1051/0004-6361/201322068}

\bibitem[{{Astropy Collaboration} {et~al.}(2018){Astropy Collaboration},
  {Price-Whelan}, {Sip{\H{o}}cz}, {G{\"u}nther}, {Lim}, {Crawford}, {Conseil},
  {Shupe}, {Craig}, {Dencheva}, {Ginsburg}, {VanderPlas}, {Bradley},
  {P{\'e}rez-Su{\'a}rez}, {de Val-Borro}, {Aldcroft}, {Cruz}, {Robitaille},
  {Tollerud}, {Ardelean}, {Babej}, {Bach}, {Bachetti}, {Bakanov}, {Bamford},
  {Barentsen}, {Barmby}, {Baumbach}, {Berry}, {Biscani}, {Boquien}, {Bostroem},
  {Bouma}, {Brammer}, {Bray}, {Breytenbach}, {Buddelmeijer}, {Burke},
  {Calderone}, {Cano Rodr{\'\i}guez}, {Cara}, {Cardoso}, {Cheedella}, {Copin},
  {Corrales}, {Crichton}, {D'Avella}, {Deil}, {Depagne}, {Dietrich}, {Donath},
  {Droettboom}, {Earl}, {Erben}, {Fabbro}, {Ferreira}, {Finethy}, {Fox},
  {Garrison}, {Gibbons}, {Goldstein}, {Gommers}, {Greco}, {Greenfield},
  {Groener}, {Grollier}, {Hagen}, {Hirst}, {Homeier}, {Horton}, {Hosseinzadeh},
  {Hu}, {Hunkeler}, {Ivezi{\'c}}, {Jain}, {Jenness}, {Kanarek}, {Kendrew},
  {Kern}, {Kerzendorf}, {Khvalko}, {King}, {Kirkby}, {Kulkarni}, {Kumar},
  {Lee}, {Lenz}, {Littlefair}, {Ma}, {Macleod}, {Mastropietro}, {McCully},
  {Montagnac}, {Morris}, {Mueller}, {Mumford}, {Muna}, {Murphy}, {Nelson},
  {Nguyen}, {Ninan}, {N{\"o}the}, {Ogaz}, {Oh}, {Parejko}, {Parley}, {Pascual},
  {Patil}, {Patil}, {Plunkett}, {Prochaska}, {Rastogi}, {Reddy Janga},
  {Sabater}, {Sakurikar}, {Seifert}, {Sherbert}, {Sherwood-Taylor}, {Shih},
  {Sick}, {Silbiger}, {Singanamalla}, {Singer}, {Sladen}, {Sooley},
  {Sornarajah}, {Streicher}, {Teuben}, {Thomas}, {Tremblay}, {Turner},
  {Terr{\'o}n}, {van Kerkwijk}, {de la Vega}, {Watkins}, {Weaver}, {Whitmore},
  {Woillez}, {Zabalza}, \& {Astropy Contributors}}]{Astropy2018}
{Astropy Collaboration}, {Price-Whelan}, A.~M., {Sip{\H{o}}cz}, B.~M., {et~al.}
  2018, \aj, 156, 123, \dodoi{10.3847/1538-3881/aabc4f}

\bibitem[{{Chevalier} \& {Ilovaisky}(1998)}]{CI1998}
{Chevalier}, C., \& {Ilovaisky}, S.~A. 1998, \aap, 330, 201.
\newblock \doarXiv{astro-ph/9710008}

\bibitem[{{Choi} {et~al.}(2016){Choi}, {Dotter}, {Conroy}, {Cantiello},
  {Paxton}, \& {Johnson}}]{Choi2016}
{Choi}, J., {Dotter}, A., {Conroy}, C., {et~al.} 2016, \apj, 823, 102,
  \dodoi{10.3847/0004-637X/823/2/102}

\bibitem[{{Dotter}(2016)}]{Dotter2016}
{Dotter}, A. 2016, \apjs, 222, 8, \dodoi{10.3847/0067-0049/222/1/8}

\bibitem[{{Fabbiano}(1989)}]{Fabbiano1989}
{Fabbiano}, G. 1989, \araa, 27, 87, \dodoi{10.1146/annurev.aa.27.090189.000511}

\bibitem[{{Fabbiano}(2006)}]{Fabbiano2006}
---. 2006, \araa, 44, 323, \dodoi{10.1146/annurev.astro.44.051905.092519}

\bibitem[{{Freeman} {et~al.}(2002){Freeman}, {Kashyap}, {Rosner}, \&
  {Lamb}}]{Freeman2002}
{Freeman}, P.~E., {Kashyap}, V., {Rosner}, R., \& {Lamb}, D.~Q. 2002, \apjs,
  138, 185, \dodoi{10.1086/324017}

\bibitem[{{Fruscione} {et~al.}(2006){Fruscione}, {McDowell}, {Allen},
  {Brickhouse}, {Burke}, {Davis}, {Durham}, {Elvis}, {Galle}, {Harris},
  {Huenemoerder}, {Houck}, {Ishibashi}, {Karovska}, {Nicastro}, {Noble},
  {Nowak}, {Primini}, {Siemiginowska}, {Smith}, \& {Wise}}]{Fruscione2006}
{Fruscione}, A., {McDowell}, J.~C., {Allen}, G.~E., {et~al.} 2006, in Society
  of Photo-Optical Instrumentation Engineers (SPIE) Conference Series, Vol.
  6270, Society of Photo-Optical Instrumentation Engineers (SPIE) Conference
  Series, ed. D.~R. {Silva} \& R.~E. {Doxsey}, 62701V,
  \dodoi{10.1117/12.671760}

\bibitem[{{Harris} {et~al.}(2020){Harris}, {Millman}, {van der Walt},
  {Gommers}, {Virtanen}, {Cournapeau}, {Wieser}, {Taylor}, {Berg}, {Smith},
  {Kern}, {Picus}, {Hoyer}, {van Kerkwijk}, {Brett}, {Haldane}, {del R{\'\i}o},
  {Wiebe}, {Peterson}, {G{\'e}rard-Marchant}, {Sheppard}, {Reddy}, {Weckesser},
  {Abbasi}, {Gohlke}, \& {Oliphant}}]{Harris2020}
{Harris}, C.~R., {Millman}, K.~J., {van der Walt}, S.~J., {et~al.} 2020, \nat,
  585, 357, \dodoi{10.1038/s41586-020-2649-2}

\bibitem[{{Jin} \& {Kong}(2019)}]{Jin2019}
{Jin}, R., \& {Kong}, A. K.~H. 2019, \apj, 879, 112,
  \dodoi{10.3847/1538-4357/ab2461}

\bibitem[{{Joye} \& {Mandel}(2003)}]{Joye2003}
{Joye}, W.~A., \& {Mandel}, E. 2003, in Astronomical Society of the Pacific
  Conference Series, Vol. 295, Astronomical Data Analysis Software and Systems
  XII, ed. H.~E. {Payne}, R.~I. {Jedrzejewski}, \& R.~N. {Hook}, 489

\bibitem[{{Kaaret} {et~al.}(2001){Kaaret}, {Prestwich}, {Zezas}, {Murray},
  {Kim}, {Kilgard}, {Schlegel}, \& {Ward}}]{Kaaret2001}
{Kaaret}, P., {Prestwich}, A.~H., {Zezas}, A., {et~al.} 2001, \mnras, 321, L29,
  \dodoi{10.1046/j.1365-8711.2001.04064.x}

\bibitem[{{Kim} {et~al.}(2007){Kim}, {Kim}, {Wilkes}, {Green}, {Kim},
  {Anderson}, {Barkhouse}, {Evans}, {Ivezi{\'c}}, {Karovska}, {Kashyap}, {Lee},
  {Maksym}, {Mossman}, {Silverman}, \& {Tananbaum}}]{Kim2007}
{Kim}, M., {Kim}, D.-W., {Wilkes}, B.~J., {et~al.} 2007, \apjs, 169, 401,
  \dodoi{10.1086/511634}

\bibitem[{{Kuntz} {et~al.}(2016){Kuntz}, {Long}, \& {Kilgard}}]{Kuntz2016}
{Kuntz}, K.~D., {Long}, K.~S., \& {Kilgard}, R.~E. 2016, \apj, 827, 46,
  \dodoi{10.3847/0004-637X/827/1/46}

\bibitem[{{Lehmer} {et~al.}(2019){Lehmer}, {Eufrasio}, {Tzanavaris},
  {Basu-Zych}, {Fragos}, {Prestwich}, {Yukita}, {Zezas}, {Hornschemeier}, \&
  {Ptak}}]{Lehmer2019}
{Lehmer}, B.~D., {Eufrasio}, R.~T., {Tzanavaris}, P., {et~al.} 2019, \apjs,
  243, 3, \dodoi{10.3847/1538-4365/ab22a8}

\bibitem[{{Luan} {et~al.}(2018){Luan}, {Jones}, {Forman}, {Bogd{\'a}n},
  {Andrade-Santos}, {Goulding}, {Hickox}, {Hou}, \& {Li}}]{Luan2018}
{Luan}, L., {Jones}, C., {Forman}, W.~R., {et~al.} 2018, \apj, 862, 73,
  \dodoi{10.3847/1538-4357/aaca94}

\bibitem[{{McBride} {et~al.}(2008){McBride}, {Coe}, {Negueruela}, {Schurch}, \&
  {McGowan}}]{McBride2008}
{McBride}, V.~A., {Coe}, M.~J., {Negueruela}, I., {Schurch}, M.~P.~E., \&
  {McGowan}, K.~E. 2008, \mnras, 388, 1198,
  \dodoi{10.1111/j.1365-2966.2008.13410.x}

\bibitem[{{McQuinn} {et~al.}(2016){McQuinn}, {Skillman}, {Dolphin}, {Berg}, \&
  {Kennicutt}}]{McQuinn2016}
{McQuinn}, K. B.~W., {Skillman}, E.~D., {Dolphin}, A.~E., {Berg}, D., \&
  {Kennicutt}, R. 2016, \apj, 826, 21, \dodoi{10.3847/0004-637X/826/1/21}

\bibitem[{{Mineo} {et~al.}(2012){Mineo}, {Gilfanov}, \& {Sunyaev}}]{Mineo2012}
{Mineo}, S., {Gilfanov}, M., \& {Sunyaev}, R. 2012, \mnras, 419, 2095,
  \dodoi{10.1111/j.1365-2966.2011.19862.x}

\bibitem[{{Mutchler} {et~al.}(2005){Mutchler}, {Beckwith}, {Bond}, {Christian},
  {Frattare}, {Hamilton}, {Hamilton}, {Levay}, {Noll}, \&
  {Royle}}]{Mutchler2005}
{Mutchler}, M., {Beckwith}, S.~V.~W., {Bond}, H., {et~al.} 2005, in American
  Astronomical Society Meeting Abstracts, Vol. 206, American Astronomical
  Society Meeting Abstracts \#206, 13.07

\bibitem[{{Paxton} {et~al.}(2011){Paxton}, {Bildsten}, {Dotter}, {Herwig},
  {Lesaffre}, \& {Timmes}}]{Paxton2011}
{Paxton}, B., {Bildsten}, L., {Dotter}, A., {et~al.} 2011, \apjs, 192, 3,
  \dodoi{10.1088/0067-0049/192/1/3}

\bibitem[{{Paxton} {et~al.}(2013){Paxton}, {Cantiello}, {Arras}, {Bildsten},
  {Brown}, {Dotter}, {Mankovich}, {Montgomery}, {Stello}, {Timmes}, \&
  {Townsend}}]{Paxton2013}
{Paxton}, B., {Cantiello}, M., {Arras}, P., {et~al.} 2013, \apjs, 208, 4,
  \dodoi{10.1088/0067-0049/208/1/4}

\bibitem[{{Paxton} {et~al.}(2015){Paxton}, {Marchant}, {Schwab}, {Bauer},
  {Bildsten}, {Cantiello}, {Dessart}, {Farmer}, {Hu}, {Langer}, {Townsend},
  {Townsley}, \& {Timmes}}]{Paxton2015}
{Paxton}, B., {Marchant}, P., {Schwab}, J., {et~al.} 2015, \apjs, 220, 15,
  \dodoi{10.1088/0067-0049/220/1/15}

\bibitem[{{Paxton} {et~al.}(2018){Paxton}, {Schwab}, {Bauer}, {Bildsten},
  {Blinnikov}, {Duffell}, {Farmer}, {Goldberg}, {Marchant}, {Sorokina},
  {Thoul}, {Townsend}, \& {Timmes}}]{Paxton2018}
{Paxton}, B., {Schwab}, J., {Bauer}, E.~B., {et~al.} 2018, \apjs, 234, 34,
  \dodoi{10.3847/1538-4365/aaa5a8}

\bibitem[{{Perez} \& {Granger}(2007)}]{Perez2007}
{Perez}, F., \& {Granger}, B.~E. 2007, Computing in Science and Engineering, 9,
  21, \dodoi{10.1109/MCSE.2007.53}

\bibitem[{{Prestwich} {et~al.}(2003){Prestwich}, {Irwin}, {Kilgard}, {Krauss},
  {Zezas}, {Primini}, {Kaaret}, \& {Boroson}}]{Prestwich2003}
{Prestwich}, A.~H., {Irwin}, J.~A., {Kilgard}, R.~E., {et~al.} 2003, \apj, 595,
  719, \dodoi{10.1086/377366}

\bibitem[{{Remillard} \& {McClintock}(2006)}]{Remillard2006}
{Remillard}, R.~A., \& {McClintock}, J.~E. 2006, \araa, 44, 49,
  \dodoi{10.1146/annurev.astro.44.051905.092532}

\bibitem[{{Sell} {et~al.}(2019){Sell}, {Zezas}, {Williams}, {Andrews},
  {Gazeas}, {Gallagher}, \& {Ptak}}]{Sell2019}
{Sell}, P.~H., {Zezas}, A., {Williams}, S.~J., {et~al.} 2019, in IAU Symposium,
  Vol. 346, IAU Symposium, ed. L.~M. {Oskinova}, E.~{Bozzo}, T.~{Bulik}, \&
  D.~R. {Gies}, 344--349, \dodoi{10.1017/S1743921318008190}

\bibitem[{{Shapiro} \& {Teukolsky}(1983)}]{Shapiro1983}
{Shapiro}, S.~L., \& {Teukolsky}, S.~A. 1983, {Black holes, white dwarfs, and
  neutron stars : the physics of compact objects}

\bibitem[{{Tauris} \& {van den Heuvel}(2006)}]{Tauris2006}
{Tauris}, T.~M., \& {van den Heuvel}, E.~P.~J. 2006, {Formation and evolution
  of compact stellar X-ray sources}, Vol.~39, 623--665

\bibitem[{{Terashima} {et~al.}(2006){Terashima}, {Inoue}, \&
  {Wilson}}]{Terashima2006}
{Terashima}, Y., {Inoue}, H., \& {Wilson}, A.~S. 2006, \apj, 645, 264,
  \dodoi{10.1086/504251}

\bibitem[{{Terashima} \& {Wilson}(2004)}]{Terashima2004}
{Terashima}, Y., \& {Wilson}, A.~S. 2004, \apj, 601, 735,
  \dodoi{10.1086/380505}

\bibitem[{{Tody}(1986)}]{Tody1986}
{Tody}, D. 1986, in Society of Photo-Optical Instrumentation Engineers (SPIE)
  Conference Series, Vol. 627, Instrumentation in astronomy VI, ed. D.~L.
  {Crawford}, 733, \dodoi{10.1117/12.968154}

\bibitem[{{Tody}(1993)}]{Tody1993}
{Tody}, D. 1993, in Astronomical Society of the Pacific Conference Series,
  Vol.~52, Astronomical Data Analysis Software and Systems II, ed. R.~J.
  {Hanisch}, R.~J.~V. {Brissenden}, \& J.~{Barnes}, 173

\bibitem[{{Virtanen} {et~al.}(2020){Virtanen}, {Gommers}, {Oliphant},
  {Haberland}, {Reddy}, {Cournapeau}, {Burovski}, {Peterson}, {Weckesser},
  {Bright}, {van der Walt}, {Brett}, {Wilson}, {Millman}, {Mayorov}, {Nelson},
  {Jones}, {Kern}, {Larson}, {Carey}, {Polat}, {Feng}, {Moore}, {VanderPlas},
  {Laxalde}, {Perktold}, {Cimrman}, {Henriksen}, {Quintero}, {Harris},
  {Archibald}, {Ribeiro}, {Pedregosa}, {van Mulbregt}, \& {SciPy 1. 0
  Contributors}}]{Virtanen2020}
{Virtanen}, P., {Gommers}, R., {Oliphant}, T.~E., {et~al.} 2020, Nature
  Methods, 17, 261, \dodoi{10.1038/s41592-019-0686-2}

\end{thebibliography}
\bibliographystyle{aasjournal}
\end{document}